\documentclass[aps,onecolumn,preprintnumbers,pra]{revtex4-2}
\usepackage{caption}
\usepackage{tabularx}
\usepackage{graphicx}  
\usepackage{subfigure}
\usepackage{multirow}
\usepackage{mathtools}
\usepackage{notes2bib}
\usepackage{tikz}

 \Large

\usepackage[utf8]{inputenc}
\usepackage[english]{babel}

\usepackage{fancyhdr}
\usepackage{parskip}
\usepackage[T1]{fontenc}
\usepackage{dcolumn}   

\usepackage{bm}        
\usepackage{amsfonts}  
\usepackage{amsmath}   
\usepackage{amssymb}   

\newcommand{\pwisein}{\left\{ \begin{array}{ll}}
\newcommand{\pwiseout}{\end{array}\right.}

\renewcommand{\det}[1]{\mathrm{det}\left( #1 \right)}

\usepackage[unicode=true,pdfusetitle,
 bookmarks=true,bookmarksnumbered=false,bookmarksopen=false,
 breaklinks=false,pdfborder={0 0 0},pdfborderstyle={},backref=false,colorlinks=true]
 {hyperref}
\hypersetup{allcolors=blue}

\makeatletter
\renewcommand\NAT@citesuper[3]{\ifNAT@swa
\unskip\hspace{1\p@}\textsuperscript{[#1]}%
\if\relax#3\relax\else\ [#3]\fi\else [#1]\fi\endgroup}
\makeatother

\usepackage [english]{babel}
\usepackage [autostyle, english = american]{csquotes}
\MakeOuterQuote{"}

\begin{document}
\setcitestyle{super}
\title{The robustness of composite pulses elucidated by classical mechanics. II. \linebreak The role of initial state imperfection}

\author{Jonathan Berkheim, David J. Tannor}

\affiliation {\it Department of Chemical and Biological Physics, Weizmann Institute of Science, 76100, Rehovot, Israel}


\begin{abstract}  
In nuclear magnetic resonance (NMR), Composite Pulses (CPs) are widely used to correct for pulse imperfections, e.g., RF field inhomogeneity and resonance offset. Although robust pulse sequences have been developed throughout the years, the imperfection of the initial state has not been widely discussed in the literature as an additional systematic error. In previous work, we developed a classical canonical framework to perform stability analysis and used this as a measure of CP robustness. In that work, a single initial condition was allowed to evolve under various pulse imperfections. The current work extends this approach to $2D$ distributions of initial conditions on the Bloch Sphere; the objective is to minimize the area in order to preserve coherence, while maximizing population inversion of the entire distribution. As a case study, we investigate Levitt's $90(x)180(y)90(x)$ pulse sequence, when there is a spread in initial conditions. The canonical framework enables us to assess the robustness of Levitt's pulse sequence, and we find that it is maintained to a great extent even when considering a spread of initial conditions. Nevertheless, by conducting a numerical optimization, we have identified several variants of Levitt's pulse sequence that produce a larger coherent population inversion when there is a spread in initial conditions.
\end{abstract}

\maketitle 
\section{Introduction: the ensemble of initial conditions}
\subsection{Main goals}
Composite Pulses (CPs) are widely employed in physical experiments that involve population inversion, e.g., in nuclear magnetic resonance (NMR), optical spectroscopy, optimal control experiments, and quantum information processing.\citep{Hahn1950SpinEchoes,Meiboom1958ModifiedTimes,Levitt1979NMRPulse,Tycko1985CompositeDistortion,Torosov2011SmoothProcessing,Genov2014CorrectionPulses,Kyoseva2019Detuning-modulatedControl}
They provide a collective correction to various pulse imperfections, e.g., RF field inhomogeneity and resonance offsets.\citep{Levitt1982SymmetricalInhomogeneity,Levitt1982SymmetricalOffset,Levitt1986CompositePulses}

In addition to pulse imperfections, there is almost always a spread in initial conditions. E.g., in solid-state NMR, structural imperfections like lattice distortions create a spatial inhomogeneity in the Hamiltonian;\citep{Haeberlen1976HighAveraging} under Boltzmann statistics, this inhomogeneity produces a spread of the initial state of the system, as $\rho_{0}\propto \exp(-H/k_{B}T)$. Moreover, the final conditions produced by an inhomogeneous Hamiltonian in one pulse segment lead to a spread of ICs for the next pulse segment.\citep{Berkheim2026TheGlobe} 
The sensitivity to IC has been addressed indirectly by various approaches. An early geometrical approach suggested CPs robust to pulse inhomogeneity, independent of IC.\citep{Tycko1985CompositeDistortion} This work was later extended to arbitrarily accurate sequences that were derived from systematic control principles.\citep{Brown2004ArbitrarilySequences} In parallel, design principles for broadband, narrowband and passband CPs were established and classified according to their robustness to pulse inhomogeneity when various initial states are taken into account.\citep{Wimperis1994BroadbandExperiments}

Despite the success of the above approaches, the imperfection of initial conditions has not been treated as a separate systematic error, in the sense of finite initial distribution on the Bloch Sphere. Within the framework of the canonical coordinates $(\phi,\eta)=(\tan^{-1}(y/x),z)$ defined in our recent work,\citep{Berkheim2026TheGlobe} the treatment of a $2D$ distribution of initial conditions is a natural extension. We will analyze the robustness of CPs under a spread of ICs, using the canonical coordinates and stability analysis. As a case study, we will examine Levitt's $90(x)180(y)90(x)$ pulse sequence. In our previous work, we demonstrated that in the case of resonance offset this pulse contracts the ensemble's effective length, implying that contraction may be observed when a spread of ICs is considered. Although area contraction is geometrically forbidden, we will show that the robustness of Levitt's pulse sequence is maintained to a great extent even if there is a spread of ICs.

Levitt's original work starts with a single IC, i.e., a $0D$ initial domain, $\textbf{r}_{0}=(0,0,1)$, that evolved under a variety of pulse imperfections to yield a $1D $ final domain. In the current work, we start with a $2D$ set of ICs distributed around $\textbf{r}_{0}=(0,0,1)$. Each of these ICs will evolve under a variety of pulse imperfections - either RF field inhomogeneity or resonance offset - to yield a $2D$ final distribution. When stacked together, the distributions form a $2D$ final domain, which is visualized by the projected area of the ensemble. The successive evolution of a $2D$ domain of ICs, each time under a different value of the imperfection, is precisely equivalent to the successive evolution of a single IC, each time under the entire range of the imperfection and then taking the union of all the evolved distributions. The former approach exploits the properties of the stability matrix, thus avoiding the averaging process considered in our previous work.\citep{Berkheim2026TheGlobe}

Robustness to ICs will be measured by a dual-scale evaluation framework: (1) a "macroscopic" approach involving a direct calculation of the projected area of the ensemble, (2) a "microscopic" approach involving a qualitative assessment of shear coefficients. The latter is based on the stability matrix elements in an extended phase space, where the imperfection is considered as an additional coordinate. The co-area formula elegantly connects these two measures.

Although Levitt's pulse sequence is generally robust in the $\eta$-direction for initial conditions, we show that it can be improved in the $\phi$-direction. To show this, we perform the same experiment as before, but scanning over a range of pulses, distinguished by different time durations $\tau$ and axes of rotation $\hat{\textbf{n}}_{k}$ of each pulse segment. We will show a few variants of Levitt's pulse sequence that improve the robustness in the $\phi$-direction with respect to ICs.
 
\subsection{The evolution of projected area}
In seminal works on CPs, only Hamiltonian imperfections are considered; here, state imperfections are also considered. In this work, we separately consider the usual pulse imperfections $w=\{\Omega_{1},\Delta\}$ - either the RF field inhomogeneity $\Omega_{1}$ or the resonance offset $\Delta$ - and, in addition, we consider a spread of ICs. As long as relaxation effects are neglected, a spread of ICs does not require a mixed state treatment, but rather can be treated using a set of uncorrelated pure states.

In Hamiltonian dynamics, Liouville's theorem states that any $2N$-dimensional phase space volume (for $N=1$, an area) is preserved under a canonical transformation.\citep{Lichtenberg1992RegularDynamics,Heller2018TheSpectroscopy} In particular, the time-evolution of an initial domain under the Hamiltonian flow is area-preserving. In our case, the area of the distribution $\rho_{\text{i}}=\rho_{\text{i}}(\phi_{\text{i}},\eta_{\text{i}};w)$ has to be preserved under the action of a CP if there is a definite RF field magnitude $\Omega_1$ and resonance offset $\Delta$. This situation was excluded from our previous work, which considered only a single condition. Mathematically, the statement of Liouville's theorem is
\begin{equation}
\label{Liouville}
\frac{d\rho(\phi,\eta;w)}{dt}=\frac{\partial \rho(\phi,\eta;w)}{\partial t}-\{H(\phi,\eta;w),\rho(\phi,\eta;w)\}=0,
\end{equation}
where $\{H,\rho\}$ are the Poisson brackets of the Hamiltonian and the distribution, which are both functions of the canonical coordinates, and the imperfection $w$ is fixed to a certain value. The preservation is guaranteed because the Hamiltonian flow is symplectic; in other words, the transformation $\mathcal{M}=\partial(\phi_{\text{f}},\eta_{\text{f}})/\partial(\phi_{\text{i}},\eta_{\text{i}})$ satisfies $\det{\mathcal{M}}=1$. Since area is preserved for any choice of $w$, the generalized volume $V_{\text{i}}\equiv\mu_{3}(\rho_{\text{i}}(\phi_{\text{i}},\eta_{\text{i}}))$ is also preserved, where $w$ is treated as an additional coordinate and $\mu_{3}$ denotes volume measure. Since Liouville's theorem (eq. \ref{Liouville}) holds independently for every slice, the volume that comprises the union of the slices must also be preserved. However, while this generalized volume is preserved, the projected area of the ensemble, as defined below, is generally not preserved.

\begin{figure}[ht!]
    \centering
    \includegraphics[width=1\linewidth]{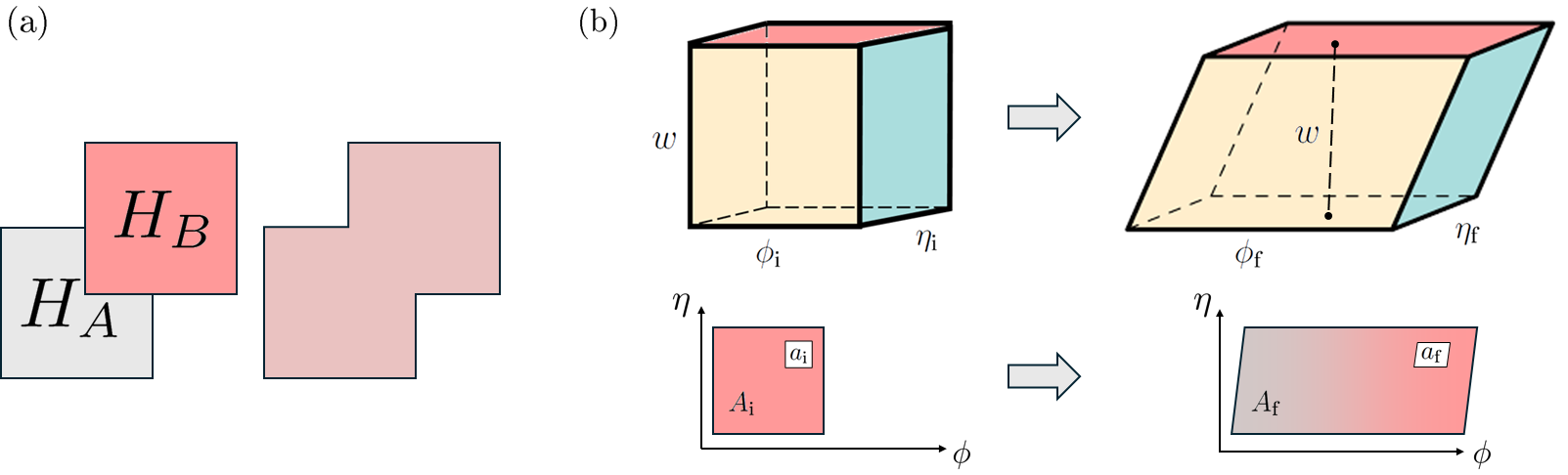}
    \caption{Preservation of volume in the extended phase space $(\phi,\eta,w)$ is not necessarily accompanied by the preservation of projected area: (a) two projections of the initial domain, each evolves under a single Hamiltonian, $H_{A}$ and $H_{B}$. The shadow area defines the union of both domains; (b) a continuous version of (a) where many sets are stacked together. The left side shows the initial domain as a cube and its projected area $A_{\text{i}}$, which consists of many square layers that perfectly overlap with each other. The right side shows the final domain as a parallelepiped, and its projected area $A_{\text{f}}$ consists of many parallelograms that do not perfectly overlap with each other, but are slightly shifted. $A_{\text{i}}$ and $A_{\text{f}}$ can be also seen as the integrations of infinitesimal areas (white-colored) $a_{\text{i}}=d\phi_{\text{i}}d\eta_{\text{i}}$ and $a_{\text{f}}=d\phi_{\text{f}}d\eta_{\text{f}}$. Overall, the projected area of the final domain is larger.}
    \label{fig:volume_transformation}
\end{figure}

The projected area $A$ of the ensemble is defined by the union of the areas spanned by the distributions associated with the set $\{w\}$, i.e.
\begin{equation}
\label{union}
A\equiv\mu_{2}
\left(\bigcup_{w\in[w_{\min},w_{\max}]} \rho(\phi,\eta;w)\right),
\end{equation}
where $\mu_{2}$ denotes area measure. This definition suffers from the fact that there is no closed functional form for the area of the union such that each point in the occupied phase space is counted only once. Figure \ref{fig:volume_transformation} illustrates how the projected area can change upon symplectic transformation of the volume element. In this work, similarly to our previous work, we consider $\Omega_{1}\in[0.8,0.9]$ and $\Delta\in[0.4,0.6]$,\citep{Levitt1979NMRPulse,Berkheim2026TheGlobe} with $11$ imperfection values in each case.

\section{Derivations and methods: a dual-scale evaluation framework}
\subsection{Macroscopic measure: ratio coefficients of the projected areas}
The macroscopic viewpoint refers to the measurement of the total $2D$ area. Instead of tracking certain trajectories, we treat the entire domain of ICs without any concern for the internal arrangement of points within it, and tackle the big-picture question: to what extent does the area get larger?

By the definition in eq. \ref{union}, the projected area at time $t$ is necessarily bigger than, or equal to, the initial projected area. We initiate all our simulations with the most compact projected area, where all initial distributions perfectly overlap with each other, such that the distributions $\rho(\phi,\eta;w)$ which are transverse cross-sections of the domain $\{\rho(\phi,\eta;w)\}_{w\in [w_{\min},w_{\max}]}$, are all identical. We will consider $\phi_{0}\in[0,2\pi]$ and $\eta_{0}\in[0.9,1]$: a rectangular distribution of ICs in the canonical framework, i.e., a dome centered at the north pole of the Bloch Sphere. We consider $200$ grid points for each coordinate.

Since all distributions are characterized by the same rectangle spanned by all points $(\phi_{0},\eta_{0})$, the projected area and the area of a single distribution are strictly equal. Any symplectic transformation applied on $\{\rho(\phi,\eta;w)\}_{w\in [w_{\min},w_{\max}]}$ will preserve or expand its projected area, so the union of projected areas can never contract with respect to the initial area at $t=0$. That being said, the following hierarchy is still possible:
\begin{equation}
\hspace{0.5cm}A(t_{1})>A_{0}; \hspace{0.5cm} A(t_{2})>A_{0}; \hspace{0.5cm} A(t_{2})<A(t_{1})
\end{equation}
such that $A(t_{2})/A(t_{1})<1$, implying a contraction along the segment $[t_{1},t_{2}]$. Still, $A(t)/A_{0}\geq 1, \forall t$, i.e., global contraction of the projected area is forbidden.

We define the ratio coefficients
\begin{equation}
\label{ratio-coeff}
R_{\text{fi}}\equiv A_{\text{f}}/A_{\text{i}},
\end{equation}
with $(\text{f},\text{i})=(0,1,2,3)$. The label $0$ designates the $t=0$ area and the labels $1,2,3$ designate the areas after the first, second, and third segments of the pulse, respectively. We aim to minimize the global expansion as expressed by the coefficient $R_{30}$. Area calculation is done numerically using the alpha-shapes technique.\citep{Edelsbrunner1983OnPlane} This calculation is controlled by the shrink factor $\alpha\in[0,1]$, where $0$ corresponds to the convex hull and $1$ corresponds to the tightest possible boundary, namely the concave hull; we have employed the default value $\alpha=0.5$. Although numerical convergence is verified, we allow a margin of error of $\pm 3\%$ in any such calculation. In particular, we claim that $R_{\text{fi}}\in[0.97,1.03]$ represents essential area preservation. We will refer to the $R_{\text{f}0} $'s as principal coefficients and to all other $R_{\text{fi}} $'s ($\text{i}\neq0$) as secondary coefficients.

\subsection{Microscopic measure: shear coefficients}
The microscopic viewpoint refers to the stability analysis of the transformation that transforms the initial domain into the final domain. Instead of tracking the entire domain of ICs as a distribution, we address the small-picture question: what is the mechanism behind the macroscopic changes of the distribution?

The projected area at the final time, $A_{\text{f}}$, spanned by all points $(\phi_{\text{f}},\eta_{\text{f}})$, manifests a union of many distributions, each generated from the ICs $(\phi_{\text{i}},\eta_{\text{i}})$ and imperfection $w$, through the transformation $u$
\begin{equation}
\label{mapping}
u:(\phi_{\text{i}},\eta_{\text{i}},w)\to(\phi_{\text{f}},\eta_{\text{f}}).
\end{equation}
Namely, as long as we treat the imperfection as an additional coordinate, the projected area is generated by a $3D$-to-$2D$ transformation; i.e., the projected area represents a reduction to a lower-dimensional space. The $2\times3$ Jacobi matrix of this transformation, denoted $\mathcal{J}$, is
\begin{equation}
\label{transformation-matrix}
\mathcal{J}(\phi,\eta,w)=\begin{pmatrix}  \partial\phi_{\text{f}}/\partial\eta_{\text{i}} && \partial\phi_{\text{f}}/\partial\phi_{\text{i}} && \partial \phi_{\text{f}}/\partial w \\ \partial\eta_{\text{f}}/\partial\eta_{\text{i}} && \partial \eta_{\text{f}}/\partial\phi_{\text{i}} && \partial \eta_{\text{f}}/\partial w 
\end{pmatrix}.
\end{equation}
Note that the left $2\times2$ submatrix is $\mathcal{M}$, the usual stability matrix. If we wish to simultaneously consider two imperfections $w_{1},w_{2}$ (e.g., $\Omega_{1}$ and $\Delta$) then $\mathcal{J}$ will become a $2\times4$ matrix. For any transformation whose matrix representation is non-square, it is common to define the local volume scaling factor,
\begin{equation}
\label{shear-coeff}
\mathcal{G}_{\text{fi}}(\phi,\eta,w)\equiv\sqrt{\det{\mathcal{J}\mathcal{J}^{T}}}=\sqrt{\sum_{\substack{I \subset \{1,2,3\} \\ |I|=2}} \det{\mathcal{J}\mathcal{J}^{T}}_{I}^{2}},
\end{equation}
where we have used the Cauchy-Binet formula in the last transition; the summation is done over the determinants of all $2\times2$ submatrices of $\mathcal{J}$. As said, the first submatrix is $\mathcal{M}$, and since $\det{\mathcal{M}}=1$, we find that $\mathcal{G}_{\text{fi}}\geq 1$ where tight equality is obtained if $\partial\phi_{\text{f}}/\partial w=\partial \eta_{\text{f}}/\partial w=0$. We denote $\mathcal{G}_{\text{fi}}$ as the shear coefficients, since they encode the microscopic, local information about the shear of volume elements in the extended phase space. These coefficients also encapsulate all inner products between the column vectors of $\mathcal{J}$. Such entities are often referred to as Gram determinants or Grammians.

The derivatives in eq. \ref{transformation-matrix} are calculated numerically using the satellite trajectory method\citep{Heller1976ClassicalDynamicsb} and it is verified that indeed $\det{\mathcal{M}}=1$ for each grid. In contrast to our previous work, satellite trajectories are not considered a separate virtual swarm, as they are already part of the examined ensemble itself. The derivatives with respect to $w$ are calculated by the symmetric differentiation rule $f'(x)\approx\frac{1}{2h}[f(x+2h)-f(x-2h)]$. Edge effects were found to contribute marginally to the distributions, and they can be neglected in the analysis of the histograms below.

It is interesting to consider the correspondence between the macroscopic and microscopic descriptions. The co-area formula, which is a generalization of Fubini's theorem and is a central result in geometrical measure theory, provides a way to associate and integral over a high-dimensional initial domain $\Omega$ of $\dim(n)$ to an integral over a lower-dimensional final domain of $\dim(k)$.\citep{Evans2025MeasureFunctions}

We begin with the initial set, $\Omega\in\mathbb{R}^{n}$. Every point $x$ has a certain weight, represented by $g(x)$. As we map these points to new points $y$ using the transformation $y=u(x)$, some regions get stretched or compressed. The Jacobian of this transformation, $|\mathcal{J}_{k}|=\sqrt{\det{\mathcal{J}\mathcal{J}^{T}}}$, acts as a scaling factor, ensuring we account for that stretching so the total density is preserved. The subscript $k$ stands for the dimensions of the final domain. Overall, the integral we are interested in is
\begin{equation}
\label{co-area-formula}
I=\int_{\Omega}g(x)|\mathcal{J}_{k}|dx.
\end{equation}
Now, instead of performing the integral globally, we look at one output value $y$ at a time. We group together all the points $x$ that are mapped to that specific $y$. This collection of points is called a slice, or fiber. Since these slices are lower-dimensional than the original space (e.g., a $1D$ curve inside a $2D$ plane), we use the Hausdorff measure, denoted $dH_{n-k}$, to designate the generalized volume of this specific slice. The subscript $n-k$ reflects the completion of dimensions of the integral.

\begin{figure}
    \centering
    \includegraphics[width=0.8\linewidth]{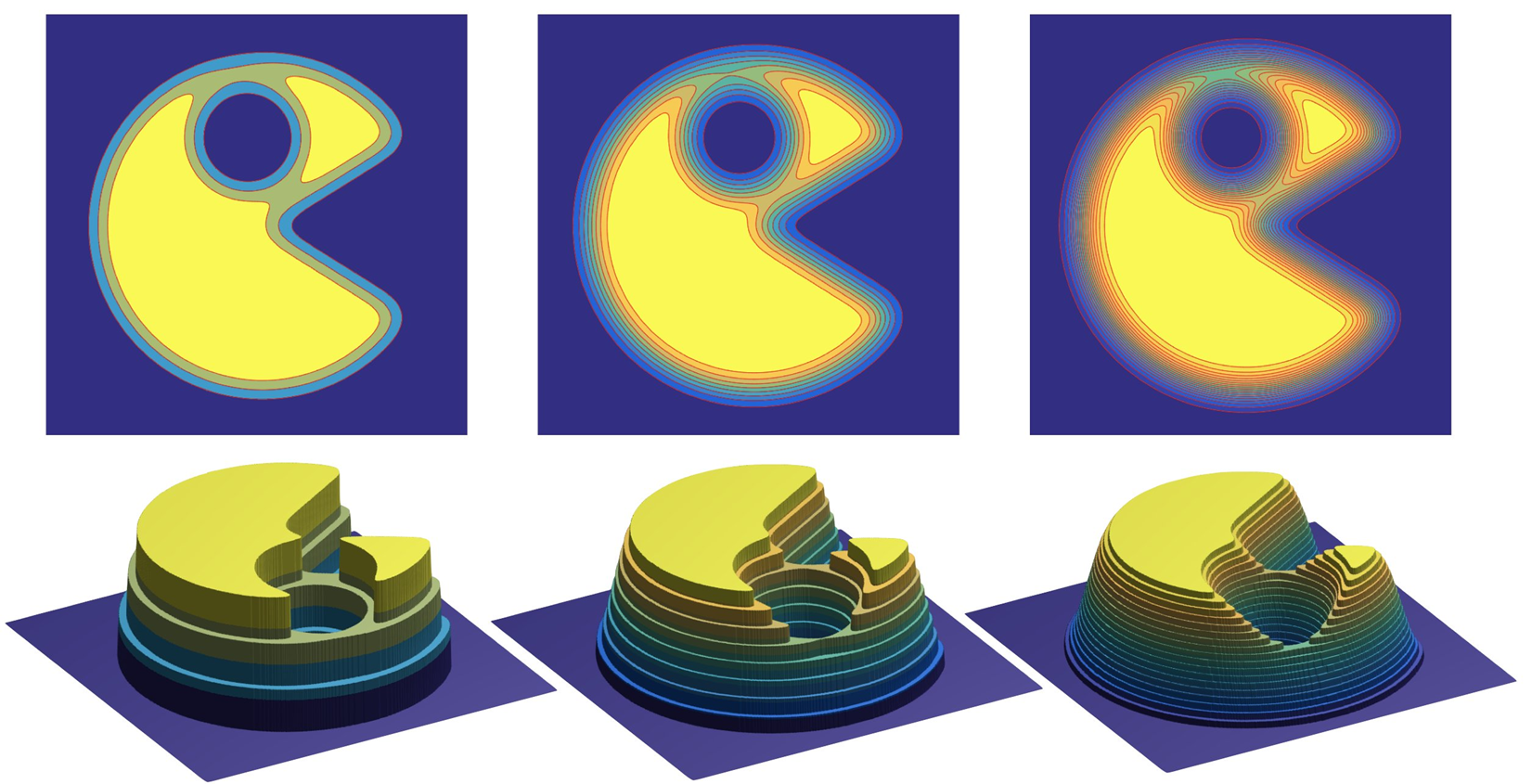}
    \caption{A visual representation of the co-area formula. The upper panel shows a scalar function $g(x)$ with $x=(x_{1},x_{2})$, where $g$ is the color scale. The lower panel shows $g(x)$ through its level sets $u(x)=y$. The total integral $\int g(x)|\mathcal{J}_{2}|dx$ is decomposed into contributions with boundary lengths $dH_{1}$ of these level sets. The sequence shows how a continuous surface emerges from an increasingly dense stack of level-set slices, i.e. the Riemann sum is refined and the sum gets closer to the continuous co-area integral.\linebreak \textit{Courtesy of Gabriel Peyré (CNRS).}} 
    \label{fig:coarea}
\end{figure}

Putting everything together, we calculate the value of $g(x)$ along each individual slice $u^{-1}(y)$ using the Hausdorff measure, then we sum those cross-sectional densities across all output values $y$ in the final space,
\begin{equation}
\label{co-area-formula-2}
I=\int_{\Omega}g(x)|\mathcal{J}_{k}|dx=\int_{\mathbb{R}^{k}}\left(\int _{u^{-1}(y)}g(x)dH_{n-k}\right)dy.
\end{equation}
Figure \ref{fig:coarea} shows a visual representation of the co-area formula.
In our system, eq. \ref{mapping} implies that $x=(\phi_{\text{i}},\eta_{\text{i}},w),n=3$ and $y=(\phi_{\text{f}},\eta_{\text{f}}), k=2$, so consequently $dH_{1}=dw$. We associate $I$ with $A_{\text{f}}=\iint_{\text{union}}{d\phi_{\text{f}}d\eta_{\text{f}}}$ (see fig. \ref{fig:volume_transformation}b) and $g$ with an inverse multiplicity function $\mathcal{N}(\phi,\eta,w)=1/\ell\leq 1$, which compensates for the multiple counting of $\ell$  overlapping layers in the union; the Jacobian is given by $\mathcal{G}_{\text{fi}}=\sqrt{\det {\mathcal J\mathcal J^{T}}}$. Inserting all these variables and functions into eq. \ref{co-area-formula}, we get
\begin{equation}
\label{co-area-applied}
A_{\text{f}}=\iint_{\text{union}}d\phi_{\text{f}}d\eta_{\text{f}}=\iint\left(\int_{u^{-1}}\mathcal{N}(\phi_{\text{f}},\eta_{\text{f}},w)dw\right)d\phi_{\text{f}}d\eta_{\text{f}}=\iiint\mathcal{N}(\phi_{\text{i}},\eta_{\text{i}},w)\sqrt{\det{\mathcal{J}\mathcal{J}^{T}}}d\phi_{\text{i}}d\eta_{\text{i}}dw,
\end{equation}
where $\sqrt{\det{\mathcal{JJ}^{T}}}$ appears due to the transformation from $da_{\text{f}}=d\phi_{\text{f}}d\eta_{\text{f}}$ to $da_{\text{i}}=d\phi_{\text{i}}d\eta_{\text{i}}$. 

From eq. \ref{co-area-applied},  the final area $A_{\text{f}}$, hence the ratio coefficient $R_{\text{fi}}=A_{\text{f}}/A_{\text{i}}$, are governed by the competition between the shear coefficients $\mathcal{G}_{\text{fi}}=\sqrt{\det{\mathcal{JJ}^{T}}}\geq1$ and the inverse multiplicity function $\mathcal{N}(\phi,\eta,w)\leq 1$. We will distinguish between three regimes based on the $\mathcal{G}_{\text{fi}}$ distributions: 1) Expansion occurs when the $\mathcal{G}_{\text{fi}}$'s approach high values that dominate over $\mathcal{N}$; 2) Preservation occurs when the $\mathcal{G}_{\text{fi}}$'s are moderately concentrated near unity, and so are the values of $\mathcal{N}$; 3) Contraction occurs when the $\mathcal{G}_{\text{fi}}$'s are sharply concentrated near unity and $\mathcal{N}$ becomes the dominant factor.

\subsection{Area optimization}
With the above tools in hand, we examine if Levitt's pulse sequence can be improved when applied to an ensemble of ICs. The ratio coefficient $R_{30}$, which measures the global change in area, is our target function; we want to minimize it so that it is as close as possible to unity. However, the preservation of the ensemble's area is not the only desired goal: we also seek a successful population inversion, which is the original purpose of CPs. The controllable parameters will be the time durations of all segments $\boldsymbol{\tau}=\{\tau_{k}\}_{k=1}^{N}$ and their axes of rotation $\underline{\underline{\hat{\textbf{n}}}}=\{\hat{\textbf{n}}_{k}\}_{k=1}^{N}$, so that we aim to solve the following optimization problem:
\begin{equation}
\label{optimization}
\left(\boldsymbol{\tau},\underline{\underline{\hat{\textbf{n}}}}\right)_{\text{opt}}=\arg\min R_{30}\left(\boldsymbol{\tau},\underline{\underline{\hat{\textbf{n}}}}\right),
\end{equation}
where, in Cartesian representation, $\hat{\textbf{n}}_{k}=(\sin\vartheta_{k}\cos\varphi_{k},\sin\vartheta_{k}\sin\varphi_{k},\cos\vartheta_{k})$. Note that the pulse imperfections and the spread of ICs are not considered control parameters at this point; they are assumed to be given. In practice, an analytic approach to eq. \ref{optimization} is doomed to fail, since $R_{30}$ is a complicated, nonlinear function of the ICs, imperfections $w$ and the control sets $\boldsymbol{\tau}$ and $\underline{\underline{\hat{\textbf{n}}}}$, which cannot be written in closed form. Instead, we perform the optimization numerically, and we restrict ourselves to 3-segment symmetric pulses of the form $\tau_{1}(\hat{\textbf{n}}_{1})\tau_{2}(\hat{\textbf{n}}_{2})\tau_{1}(\hat{\textbf{n}}_{1})$.

Each $\hat{\textbf{n}}_{k}$ will be determined by $\vartheta_{k},\varphi_{k}$, where\citep{Levitt1986CompositePulses}
\begin{equation}
\label{nominal-theta}
\vartheta_{k}=\tan^{-1}(\Omega_{1}/\Delta),
\end{equation}
and $\varphi_{k}$ is predetermined by the investigator by changing the phase of the RF transmitter. In Levitt's pulse sequence, as long as the RF is in resonance with the Larmor frequency $(\Delta=0)$, $\hat{\textbf{n}}_{k}$ is embedded in the equatorial plane. We will artificially add a free parameter $\Delta\vartheta_{k}$ to allow each $\hat{\textbf{n}}_{k}$ the ability to incline even if the system is in resonance, so that $\vartheta_{k}=\vartheta_{k}^{(0)}+\Delta\vartheta_{k}$ with $\vartheta^{(0)}_{k}$ is determined by eq. \ref{nominal-theta}. For simplicity of notation we will henceforth write $\Delta\varphi_{k}$ instead of $\varphi_{k}$, so the modified axes of rotations now read
\begin{equation}
\label{axis-of-rotation}
\hat{\textbf{n}}_{k}=\left(\sin(\vartheta^{(0)}_{k}+\Delta\vartheta_{k})\cos\Delta\varphi_{k},\sin(\vartheta^{(0)}_{k}+\Delta\vartheta_{k})\sin\Delta\varphi_{k},\cos(\vartheta^{(0)}_{k}+\Delta\vartheta_{k})\right).
\end{equation}
Evidently, the full control parameter space is $6D$, an unnecessarily big parameter space to handle and analyze. We will make some simplifying assumptions: (1) we arbitrarily fix the first axis of rotation $\hat{\textbf{n}}_{1}=\hat{\textbf{x}}$ ($\vartheta_{1}^{(0)}=\pi/2,\Delta\vartheta_{1}=0,\Delta\varphi_{1}=0$) and (2) assume the total pulse length $T$ is kept constant, namely $2\tau_{1}+\tau_{2}=T$. These simplifications enable us to scan the $3D$ parameter space $(\Delta\vartheta_{2},\Delta\varphi_{2},\tau_{1})$ in an attempt to find pulses with better $R_{30}$ compared to Levitt. To quantify whether the population inversion is comparable to Levitt's original pulse, we define the mean terminal population $\bar\eta_{3}$, which is measured at the end of the third segment $(t=T)$,
\begin{equation}
\label{eta-COM}
\bar{\eta}_3\equiv\frac{\sum_{j}\eta_{3}^{(j)}}{||\eta_{3}||}.
\end{equation}
Here, $j$ enumerates the members of the terminal set $\eta_{3}$, and $||\eta_{3}||$ is the cardinality of this set, i.e., the total number of points that form the distribution. In our simulations, the cardinality will be $200\times200\times11=440,000$. Successful population inversion is reflected by $\bar{\eta}_{3}\to-1$.

\section{Numerical results and analysis}
\subsection{The ensemble of RF field inhomogeneity}
\subsubsection{Ratio coefficients}
\begin{figure}[ht!]   
    \centering
    \includegraphics[width=0.5\linewidth]{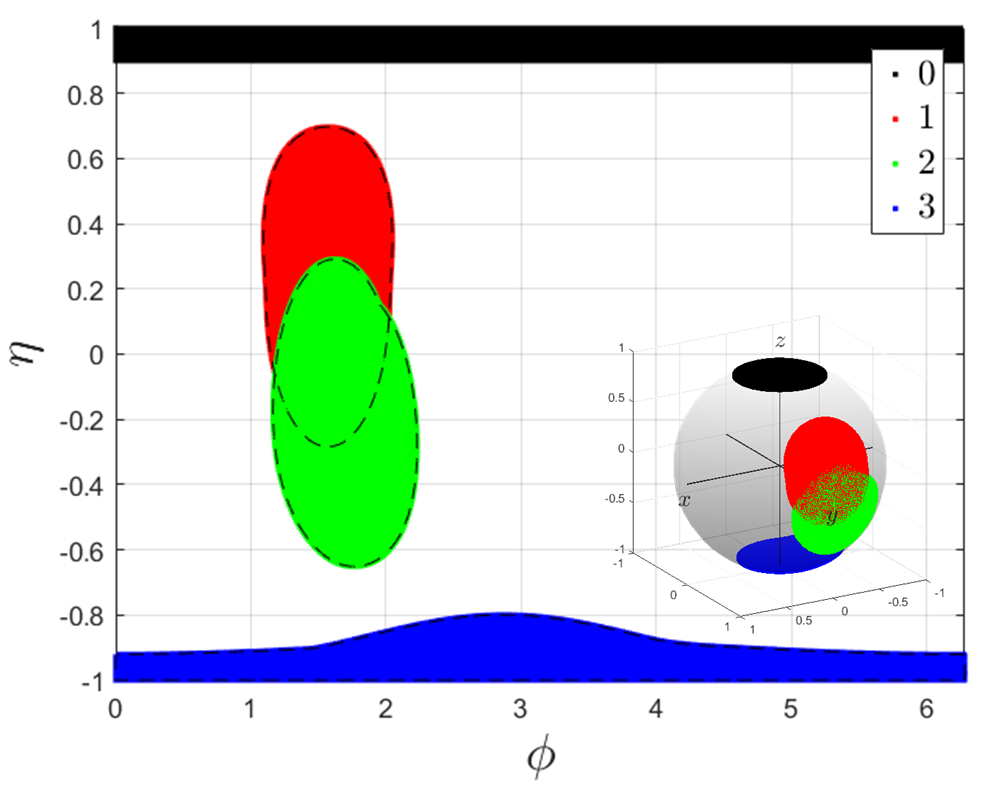}
    \caption{The time-evolution of the projected area for the ensemble of RF field inhomogeneity. The initial distribution is the black strip at the top. The first pulse segment defocuses the distribution in the $\eta$-direction and focuses it in the $\phi$-direction; the second pulse segment primarily changes the center without significant reshaping; the third pulse segment refocuses the distribution in the $\eta$-direction and defocuses it in the $\phi$-direction. The overall effect on a single IC is largely preserved when applied to the entire spread of ICs. The dashed lines mark the alpha shapes for each distribution. Inset: the same distributions presented on the Bloch Sphere.}
    \label{fig:field_ih_Rfi}
\end{figure}

Figure \ref{fig:field_ih_Rfi} shows the $2D$ distribution for the ensemble of RF field inhomogeneity in Levitt's $90(x)180(y)90(x)$ pulse sequence at four different times, both in terms of the canonical coordinates $(\phi,\eta)$ and on the Bloch Sphere. While the initial distribution $(t=0)$ is spread over a narrow band of $\eta$ values and all $\phi$ values, by the end of the first segment $(t=T/4)$, the area is stretched in the $\eta$-direction and much more localized in the $\phi$-direction. The same is true at the end of the second segment ($t=3T/4$), with the center of the distribution slightly shifted in phase space. By the end of the third segment ($t=T$), the distribution is spread over all $\phi$ values and a relatively narrow band of $\eta$ values, although the bandwidth is uneven, and broader around $\phi=0$.

Overall, Levitt's pulse sequence is robust in the $\eta$-direction and does not affect the $\phi$-direction, for a spread of ICs that evolves under RF field inhomogeneity. The ratio coefficients are tabulated in Table \ref{tab:field_ih_Rfi_table}, where it is seen that the principal coefficients $R_{\text{f}0}\geq1.20$, implying considerable area expansion; the secondary coefficients are approximately unity, implying area preservation. In the case of RF field inhomogeneity, the quasi-unitarity of Levitt's pulse sequence with respect to the evolving distribution is maintained in the second and third segments. This finding extends our previous work,\citep{Berkheim2026TheGlobe} where we demonstrated both analytically and numerically that the standard deviation of the distribution remains almost invariant along the second and third segments (when a single IC was considered). In the present case, $\bar\eta_{3}=-0.94$, indicating a relatively high efficiency of Levitt's pulse sequence in population inversion, even with a spread in initial conditions. This is to be compared to $\bar\eta_{3}=-0.99$ if only $\textbf{r}_{0}=(0,0,1)^{T}$ is considered.

    \begin{table}[ht!]   
\centering
\begin{tabular}{|c|c|c|}\hline    
\textbf{Ratio coefficient}& \textbf{Value} &\textbf{Interpretation}\\\hline
$R_{10}$& 1.21&Expansion\\\hline
$R_{21}$& 1.02&Preservation\\\hline
$R_{32}$& 0.97&Preservation\\\hline
$R_{20}$& 1.26&Expansion\\\hline
$R_{31}$& 0.97&Preservation\\\hline
$R_{30}$& 1.20&Global expansion\\ \hline
\end{tabular}
     \caption{The ratio coefficients $R_{\text{fi}}$ and their physical interpretation. Overall, at $t=T$ the area expands by $20\%$ with respect to $t=0$.}
     \label{tab:field_ih_Rfi_table}
\end{table}

\subsubsection{Shear coefficients}
\begin{figure}[ht!]
    \centering
    \includegraphics[width=0.8\linewidth]{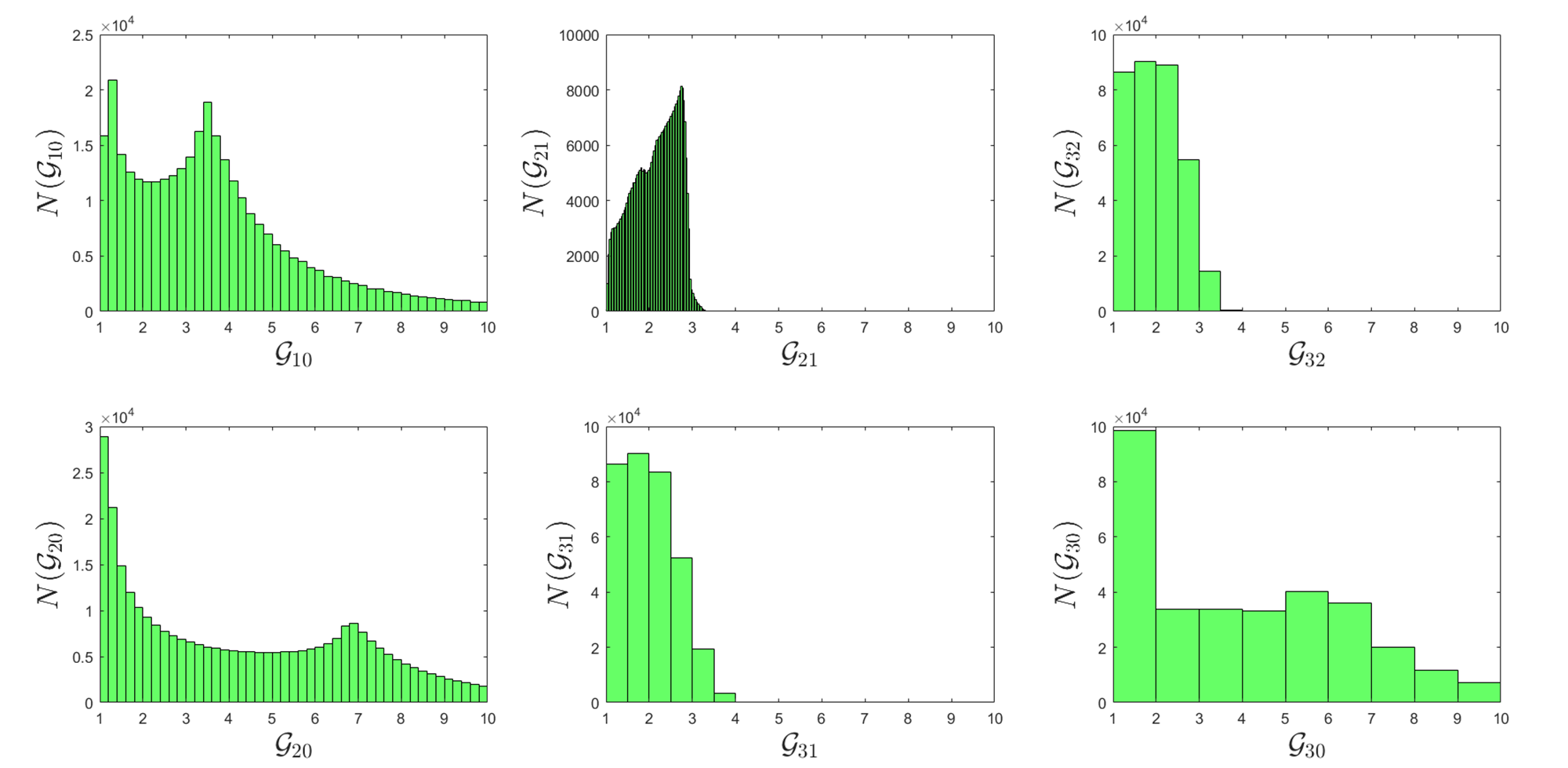}
    \caption{The distributions of shear coefficients $\mathcal{G}_{\text{fi}}$ for the ensemble of RF field inhomogeneity.}
    \label{fig:field_ih_Gfi}
\end{figure}

Figure \ref{fig:field_ih_Gfi} shows the distributions of the shear coefficients $\mathcal{G}_{\text{fi}}$, depicted as histograms $N
(\mathcal{G}_{\text{fi}})$. The shape of each distribution is tabulated in Table \ref{tab:field_ih_Gfi_table}. All distributions associated with the principal coefficients $\mathcal{G}_{\text{f0}}$ are shifted to high values, which we interpret as area expansion. In contrast, the secondary coefficients are moderately concentrated near unity, which we associate with area preservation. These observations fully agree with those of the ratio coefficients $R_{\text{fi}}$ in the previous subsection; the co-area formula, eq. \ref{co-area-applied}, explains the correspondence.

\begin{table}[ht!]   
\centering

\begin{tabular}{|c|c|c|}\hline    
\textbf{Shear coefficient}& \textbf{Shape}&\textbf{Interpretation}\\\hline
$\mathcal{G}_{10}$& Shifted to high values&Expansion\\\hline
$\mathcal{G}_{21}$& Moderately concentrated near unity&Preservation\\\hline
$\mathcal{G}_{32}$& Moderately concentrated near unity&Preservation\\\hline
$\mathcal{G}_{20}$& Shifted to high values&Expansion\\\hline
$\mathcal{G}_{31}$& Moderately concentrated near unity&Preservation\\\hline
$\mathcal{G}_{30}$& Shifted to high values&Global expansion\\ \hline

\end{tabular}
\caption{The shear coefficients $\mathcal{G}_{\text{fi}}$ and their qualitative interpretation.}
\label{tab:field_ih_Gfi_table}
\end{table}

\subsubsection{Area optimization}
\begin{figure}
    \centering
    \includegraphics[width=1\linewidth]{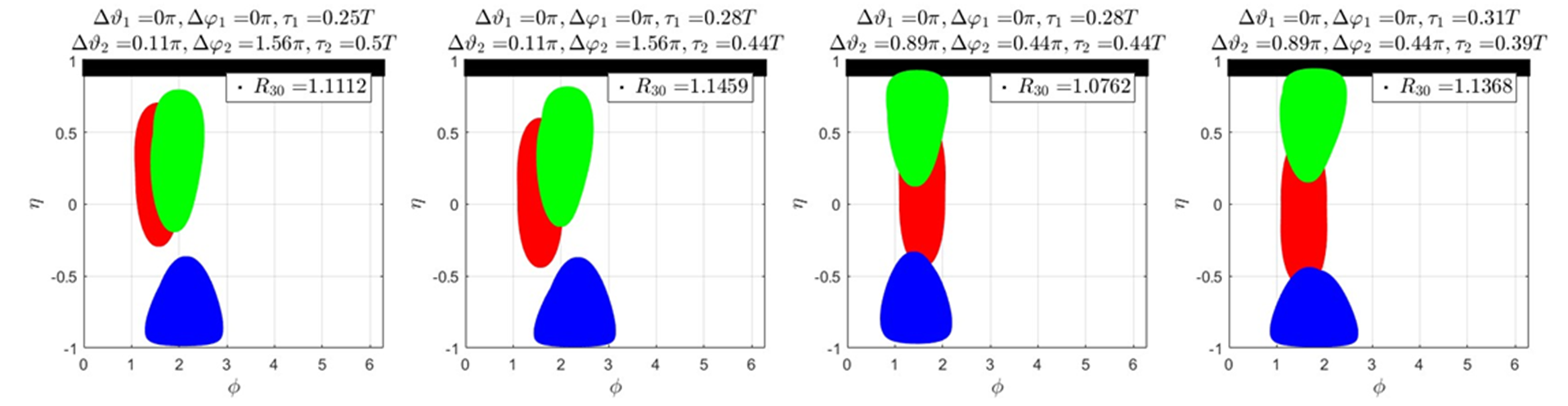}
    \caption{Four sets of parameters from the $3D$ numerical scan that outperform Levitt's pulse sequence for an ensemble of ICs subject to RF field inhomogeneity. The subfigures display the time-evolution of the ensemble at $t=0,\tau_{1},\tau_{1}+\tau_{2},T$ with the same color-coding as shown before. The optimized parameters, as detailed above each subfigure, lead to a ratio coefficient $R_{30}$ smaller than with Levitt's pulse sequence.}
    \label{fig:field_ih_area_optimization}
\end{figure}

The $3D$ scan over the parameters $(\Delta\vartheta_{2},\Delta\varphi_{2},\tau_{2})$ leads to some considerable enhancements with respect to Levitt's pulse sequence. Not surprisingly, the introduction of a $z$-component to $\boldsymbol{\Omega}$ leads to a situation resembling the ensemble of resonance offset; this brings us back to our previous work,\citep{Berkheim2026TheGlobe} where the principal difference between the ensembles was reflected in the refocusing in the $\phi$-direction. Figure \ref{fig:field_ih_area_optimization} shows the time-evolution of the initial distribution under four pulse sequences with different $(\boldsymbol{\tau},\hat{\textbf{n}})$ (see eq. \ref{optimization}) that outperform Levitt's pulse sequence for $R_{30}$ with $\bar{\eta}_{3}$ comparable to Levitt's. 

At the minimum, we obtain $R_{30}=1.08$, which is comparable to Levitt's performance in the case of resonance offset with the same parameters (see the next subsection). The structure of Levitt's pulse sequence is sometimes preserved, with $\hat{\textbf{n}}_{2}$ remaining close to the equator; however, the relative angle $\Delta\varphi_{2}-\Delta\varphi_{1}$ is smaller ($0.44\pi$ instead of $0.5\pi$). 

\subsection{The ensemble of resonance offset}
\subsubsection{Ratio coefficients}
\begin{figure}[ht!]
    \centering
    \includegraphics[width=0.5\linewidth]{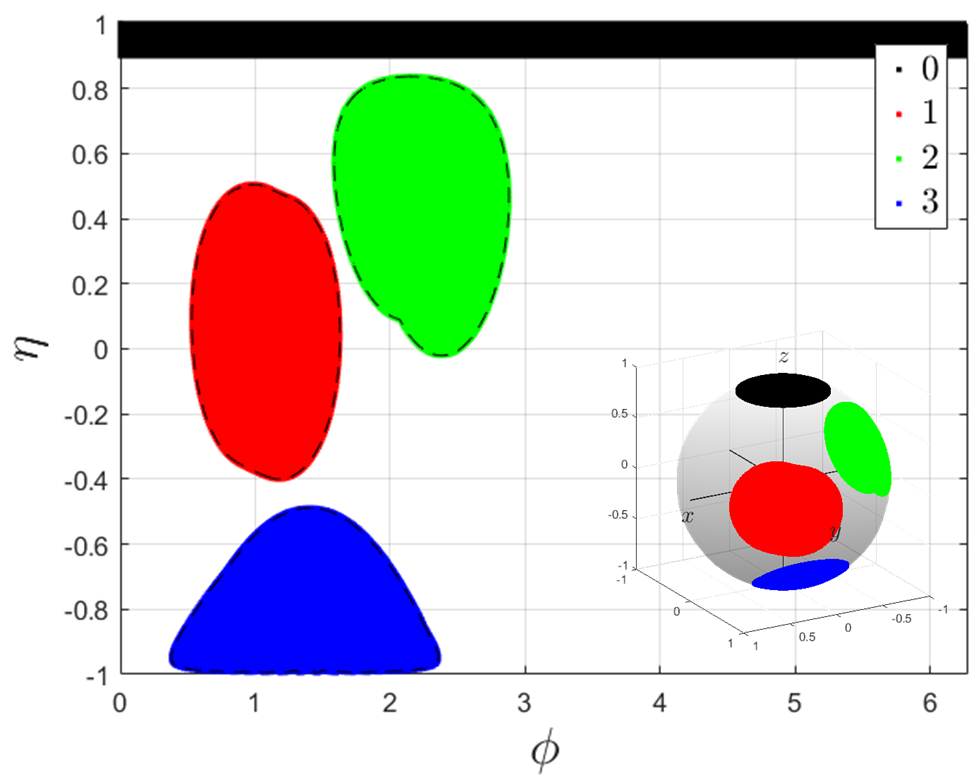}
    \caption{The time-evolution of the projected area for the ensemble of resonance offset. The initial distribution is the black strip at the top. The first segment defocuses the distribution in the $\eta$-direction and focuses it in the $\phi$-direction; the second segment changes the center without significant reshaping; the third segment slightly refocuses the distribution in the $\eta$-direction and slightly defocuses it in the $\phi$-direction. The overall effect on a single IC is largely preserved when applied to the entire spread of ICs. The dashed lines mark the alpha shapes for each distribution. Inset: the same distributions presented on the Bloch Sphere.}
    \label{fig:resonance_off_Rfi}
\end{figure}

Figure \ref{fig:resonance_off_Rfi} shows the $2D$ distribution for the ensemble of resonance offset in Levitt's $90(x)180(y)90(x)$ pulse sequence at four different times, both in terms of the canonical coordinates $(\phi,\eta)$ and on the Bloch Sphere. While the initial distribution ($t=0$) is spread over a narrow band of $\eta$ values and all $\phi$ values, by the end of the first segment $(t=T/4)$, the area is stretched in the $\eta$-direction and much more localized in the $\phi$-direction. The same is true at the end of the second segment ($t=3T/4$), with the center of the distribution slightly shifted in phase space. By the end of the third segment ($t=T$), the distribution is spread over a narrow region of $\phi$ values and a relatively broad region of $\eta$ values, such that the distribution resembles a triangle with rounded edges. Overall, Levitt's pulse sequence is robust in both the $\eta$- and $\phi$-directions, for a spread of ICs that evolves under resonance offset. The ratio coefficients are tabulated in Table \ref{tab:resonance_off_Rfi_table}, where the principal coefficients $R_{\text{f}0}>1$ (and particularly, $R_{\text{f}0}\geq1.27$ in two cases), implying considerable area expansion; two of the secondary coefficients, $R_{31}$ and $R_{32}$, are much smaller than unity, implying area contraction, and in another case, $R_{21}$, a mild expansion is observed. In the case of resonance offset, Levitt's pulse sequence is non-unitary with respect to the evolving distribution, which alternately shrinks or inflates. This finding extends our claim from the previous work,\citep{Berkheim2026TheGlobe} where we demonstrated both analytically and numerically that the standard deviation of the distribution changes significantly with each new pulse in the sequence (when a single IC was considered).

For an ensemble of initial conditions, Levitt's pulse sequence is robust in the $\phi$-direction but not so robust in the $\eta$-direction; in particular, there is a spread in $\eta$, but many ICs approach the antipode $\textbf{r}=(0,0,-1)^{T}$. Note that in Levitt's original work, under none of the pulse imperfections, $\Delta$, was the antipode reached. Globally, the change in area is almost ideal ($R_{30}\gtrsim1$). In this case, $\bar\eta_{3}=-0.79$, reflecting a slight deterioration in the performance of Levitt's pulse sequence when population inversion of the ensemble is considered; compare this to $\bar\eta_{3}=-0.85$ if only $\textbf{r}_{0}=(0,0,1)^{T}$ is considered. Levitt was aware of this suboptimal performance and suggested a 5-segment pulse that outperformed his original sequence.\citep{Levitt1982SymmetricalOffset}

\begin{table}[ht!]   
\centering

\begin{tabular}{|c|c|c|}\hline    
\textbf{Ratio coefficient}& \textbf{Value} &\textbf{Interpretation}\\\hline
$R_{10}$& 1.27&Expansion\\\hline
$R_{21}$& 1.08&Expansion\\\hline
$R_{32}$& 0.79&Contraction\\\hline
$R_{20}$& 1.37&Expansion\\\hline
$R_{31}$& 0.85&Contraction\\\hline
$R_{30}$& 1.08&Global expansion\\ \hline

\end{tabular}
\caption{The ratio coefficients $R_{\text{fi}}$ and their physical interpretation. Overall, at $t=T$ the area expands by $8\%$ with respect to $t=0$.}
\label{tab:resonance_off_Rfi_table}
\end{table}

\subsubsection{Shear coefficients}
\begin{figure}[ht!]
    \centering
    \includegraphics[width=0.8\linewidth]{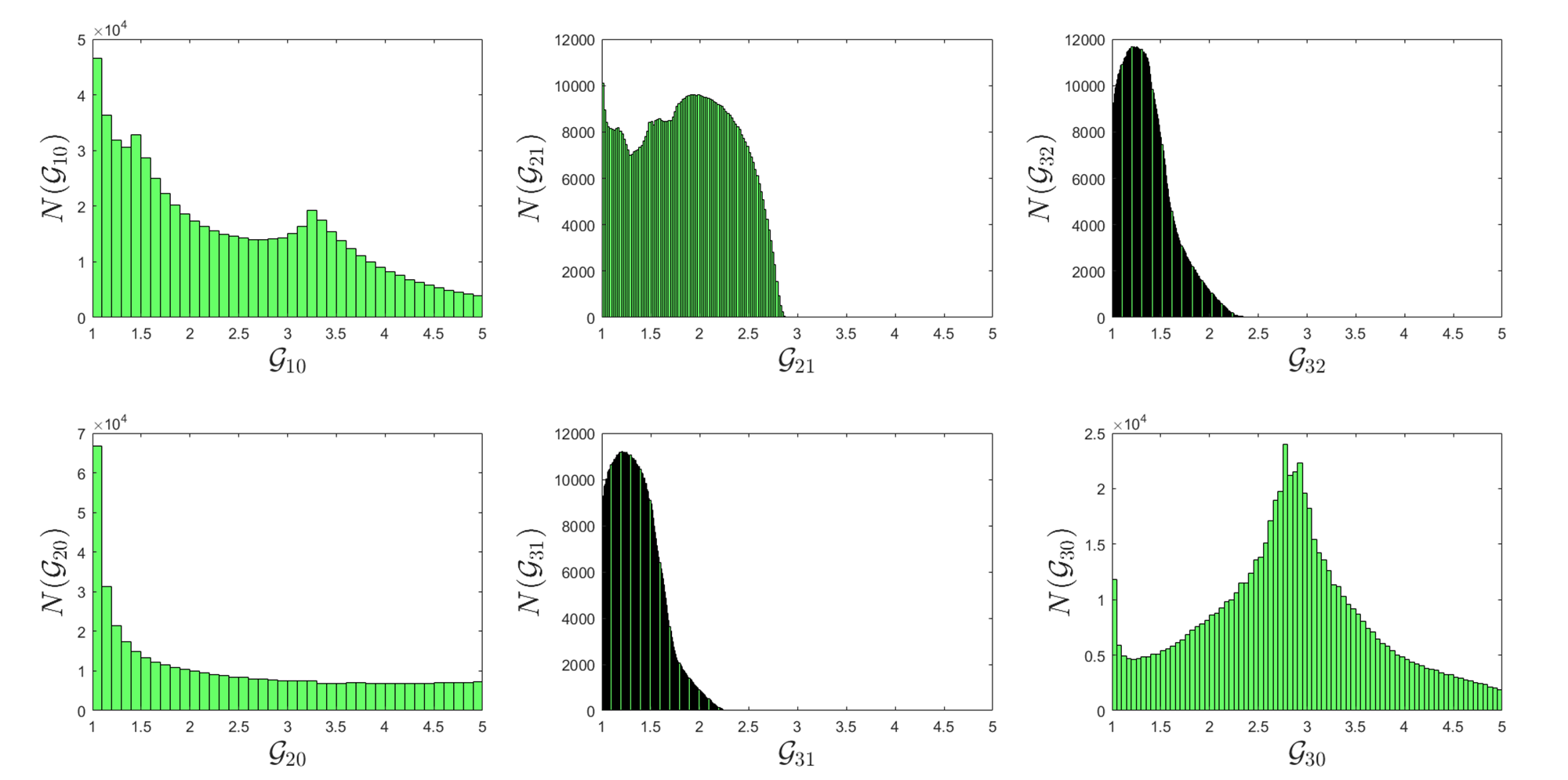}
    \caption{The distributions of shear coefficients $\mathcal{G}_{\text{fi}}$ for the ensemble of resonance offset.}
    \label{fig:resonance_off_Gfi}
\end{figure}

Figure \ref{fig:resonance_off_Gfi} shows the distributions of the shear coefficients $\mathcal{G}_{\text{fi}}$, depicted as histograms $N(\mathcal{G}_{\text{fi}})$. The shape of each distribution is tabulated in Table \ref{tab:resonance_off_Gfi_table}. All distributions associated with the principal coefficients $\mathcal{G}_{\text{f0}}$ are shifted to high values, which we interpret as area expansion. The distributions of the secondary coefficients $\mathcal{G}_{\text{32}}$ and $\mathcal{G}_{\text{31}}$ are sharply concentrated near unity, which we recognize as area contraction. The distribution of the secondary coefficient $\mathcal{G}_{\text{21}}$is slightly shifted to high values, which we associate with mild area expansion. These observations fully agree with those of the ratio coefficients $R_{\text{fi}}$ in the previous subsection; the co-area formula, eq. \ref{co-area-applied}, explains the correspondence.

\begin{table}[ht!]   
\centering

\begin{tabular}{|c|c|c|}\hline    
\textbf{Shear coefficient}& \textbf{Shape}&\textbf{Interpretation}\\\hline
$\mathcal{G}_{10}$& Shifted to high values&Expansion\\\hline
$\mathcal{G}_{21}$& Shifted to high values&Expansion\\\hline
$\mathcal{G}_{32}$& Sharply concentrated near unity&Contraction\\\hline
$\mathcal{G}_{20}$& Shifted to high values&Expansion\\\hline
$\mathcal{G}_{31}$& Sharply concentrated near unity&Contraction\\\hline
$\mathcal{G}_{30}$& Shifted to high values&Global expansion\\ \hline

\end{tabular}
\caption{The shear coefficients $\mathcal{G}_{\text{fi}}$ and their immediate interpretation.}
\label{tab:resonance_off_Gfi_table}
\end{table}

\subsubsection{Area optimization}
\begin{figure}
    \centering
    \includegraphics[width=0.8\linewidth]{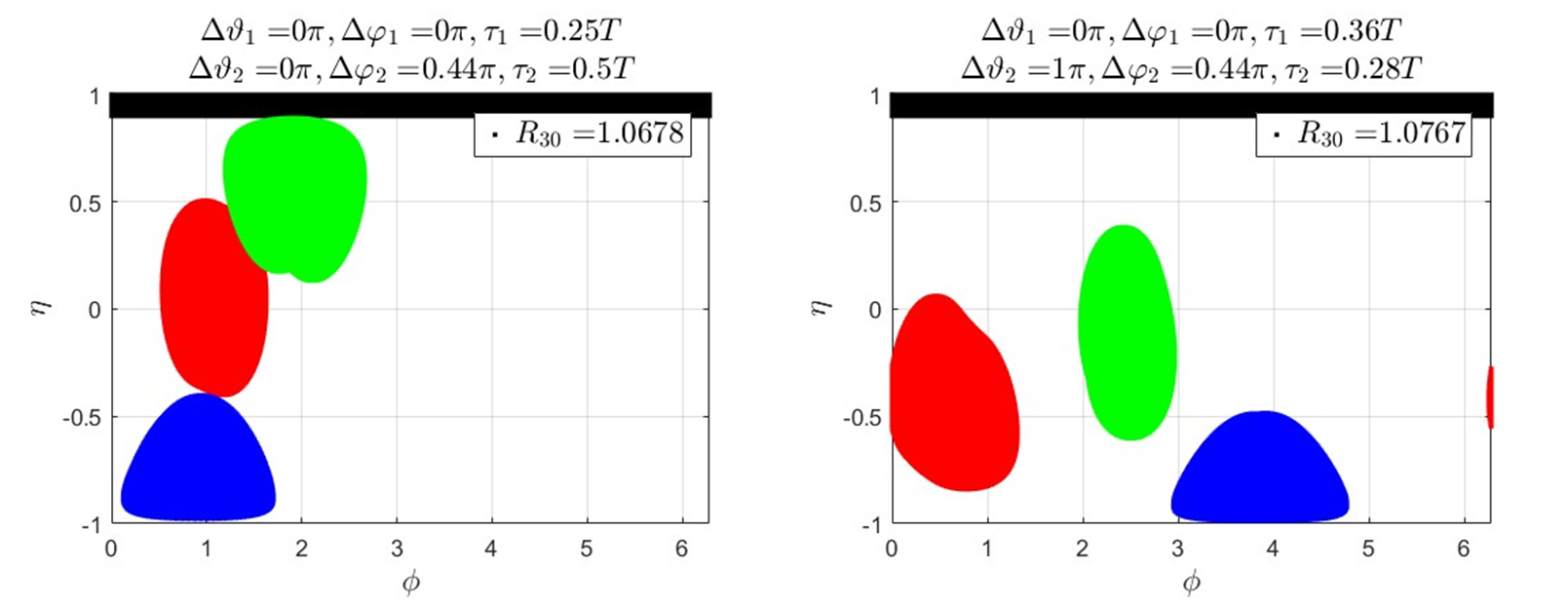}
    \caption{Two sets of parameters from the $3D$ numerical scan that lead to expansion comparable to Levitt's pulse sequence for the ensemble of ICs subject to resonance offset. The subfigures display the time-evolution of the ensemble at $t=0,\tau_{1},\tau_{1}+\tau_{2},T$ with the same color-coding as shown before. The optimized parameters, as detailed above each subfigure, lead to marginal improvement in the ratio coefficient $R_{30}$.}
    \label{fig:resonance_off_area_optimization}
\end{figure}

In view of the relatively small ratio coefficient $R_{30}=1.08$ in Levitt's pulse sequence, there is not much room to do better. Figure \ref{fig:resonance_off_area_optimization} shows the outcomes of the $3D$ scan over $(\Delta\vartheta_{2},\Delta\varphi_{2},\tau_{2})$, where it is evident that the optimized pulse sequences do not introduce any enhancement beyond the range of numerical error. Therefore, we conclude that Levitt's pulse sequence is a reasonable choice for the ensemble of ICs subject to resonance offset.

\section{Conclusion}
In this work, we introduced initial state imperfection as a systematic error that can accompany familiar pulse imperfections. While our previous work focused on the rigorous formulation of a canonical framework and an associated stability analysis, in this work we demonstrated the natural extension of these tools to distributions of ICs. Since global area contraction is forbidden under Hamiltonian flow, even in the presence of pulse imperfections, the robustness is redefined as the ability to preserve the projected area. 

We formulated a dual-scale evaluation framework for assessing the robustness of CPs within an ensemble of ICs. Beyond the direct area calculations, we found that expansion, preservation, and contraction can all be understood through the behavior of shear coefficients that stem from stability analysis. The trends of the shear coefficients correspond to the trends of the ratio coefficients that characterize the direct area calculations, as they are connected through the co-area formula. It seems fortuitous that Levitt designed a pulse sequence whose robustness is maintained even if an ensemble of ICs is considered. Now we have geometrical support for this strikingly efficient pulse sequence, though in the case of field inhomogeneity we can make the pulse sequence somewhat more robust by adding a field in the $z$-direction.

Although we focused here on the $90(x)180(y)90(x)$ pulse sequence, our method can be applied to any pulse sequence, without making assumptions about the ensemble structure or pulse form. For example, we successfully checked our method on Tycko's $180(0)180(120)180(0)$ pulse sequence,\citep{Tycko1985CompositeDistortion} claimed to be the first CP robust to RF field inhomogeneity for any IC; we found that it outperforms Levitt's pulse sequence ($R_{30}=1.05$, $\bar{\eta}_{3}=-0.95$).

\section{Acknowledgements}
J. B. dedicates this work in the memory of his friend and mentor, Prof. Issac P. Witz (1934-2026), a world-renowned cancer researcher who passed away during the preparation of the manuscript.
J. B. and D. J. T. thank Ilya Kuprov for helpful discussions. Financial support for this work was provided by the German-Israeli Foundation for Scientific Research and Development (GIF), the Israel Science Foundation (Grant Nos. 1094/16 and 1404/21), and the historic generosity of the Harold Perlman family.

\bibliography{references}

\end{document}